\documentclass[11pt]{article}

\usepackage{amsmath}
\usepackage{amssymb}
\usepackage{hyperref}
\usepackage[dvips]{graphicx}
\usepackage{cite}
\usepackage[dvips]{color} 

\usepackage{setspace} 

\usepackage[labelfont=bf,labelsep=period,justification=raggedright]{caption}

\makeatletter
\renewcommand{\@biblabel}[1]{\quad#1.}
\makeatother

\date{}

\newcommand{\Ffac}[1]{\frac{f_{#1} \rho_{#1}}{\phi}}
\newcommand{\hf}[1]{\hat{f}_{#1}}

\begin{document}

\begin{flushleft}
{\Large
\textbf{Evolution in a changing environment}
}
\\
Andrea Baronchelli$^{1}$, 
Nick Chater$^{2}$, 
Morten H. Christiansen$^{3}$,
Romualdo Pastor-Satorras$^{4, \ast}$
\\
\bf{1} Laboratory for the Modeling of Biological and Socio-technical Systems, Northeastern University, Boston MA 02115, USA\\
\bf{2} Behavioural Science Group, Warwick Business School, University of Warwick, U.K.
\\
\bf{3} Department of Psychology, Cornell University, USA, and Santa
  Fe Institute, USA
\\
\bf{4} Departament de Fisica i Enginyeria Nuclear, Universitat Politecnica de Catalunya, Campus Nord B4, 08034 Barcelona, Spain
\\
$\ast$ E-mail: romualdo.pastor@upc.edu
\end{flushleft}

\section*{Abstract}

We propose a simple model for genetic adaptation to a changing
environment, describing a fitness landscape characterized by two
maxima. One is associated with ``specialist'' individuals that are adapted to
the environment; this maximum moves over time as the environment changes.  The other maximum is static, and
represents ``generalist'' individuals not affected by environmental
changes. The rest of the landscape is occupied by ``maladapted''
individuals. Our analysis considers the evolution of these three subpopulations. Our main result is that, in presence of a sufficiently
stable environmental feature, as in the case of an unchanging aspect of
a physical habitat, specialists can dominate the population. By contrast,
rapidly changing environmental features, such as language or cultural
habits, are a moving target for the genes; here, generalists dominate, because the best
evolutionary strategy is to adopt neutral alleles not
specialized for any specific environment. The model we propose is based on simple assumptions about evolutionary dynamics and describes all possible
scenarios in a non-trivial phase diagram. The approach provides a general
framework to address such fundamental issues as the Baldwin effect,
the biological basis for language, or the ecological consequences of a
rapid climate change.


\section*{Introduction}

The issue of the evolution and adaptation in a changing environment
has recently become hotly debated in the context of climate change and
its effects on the extinction rates and the alteration of the
distribution of species
\cite{Pounds:1999fk,Parmesan:2003uq,thomas2004extinction,Sinervo:2010fk},
but it is crucial in many domains.  A classical example is given by
lactose tolerance, where the advent of dairying created an
environmental pressure in favor of this genetic trait, that in turn
further increased the benefit of dairying, in a positive feedback loop
\cite{beja2003gene}. Furthermore, it has been argued that this
so-called Baldwin effect \cite{baldwin1896new} may apply to many other
aspects of the human evolution, such as the evolution of a language
faculty. Here, an established linguistic convention would create a
selective pressure, enhancing the reproductive fitness of those
individuals that happen by chance to learn it faster or better. Over
time, less environmental exposure would therefore be needed and what was
originally a linguistic convention would eventually become encoded in the genes of
the whole population \cite{pinker2003LE}. On the other hand, it has been argued that language is a 'moving target' which changes too
rapidly for genes to follow \cite{christiansen2008language}. This paper provides an analytical framework for addressing these issues.

The study of evolution and adaptation in a
fluctuating environment from the perspective of population genetics
has been hampered by the mathematical difficulty of the problem
\cite{gillespie1994causes}. Various modeling attempts have focused on
the conditions leading to adaptation, or lack thereof, from the
perspective of the Baldwin effect
(e.g. \cite{ancel1999quantitative,chater2009restrictions}), or on
species distributions in the presence of environmental stresses
(e.g. \cite{lynchplankton1991,burgerlynchchangingeviron_1995,10.1371/journal.pbio.1000357}),
but a general picture is still lacking.  These models, in fact, while
useful and interesting, are usually defined in terms of a
large number of parameters, which compromise the possibility of
studying them thoroughly from an analytical point of view, or of
achieving fundamental insights into the problem at hand. Other work
has instead followed the quasi-species approach
\cite{PhysRevLett.84.191,springerlink:10.1006/bulm.2002.0314,PhysRevE.80.061910,PhysRevE.82.021904},
but also these models are in general characterized by considerable
mathematical complexity.

Here we propose a stochastic interacting particle model that
captures the most basic features characterizing evolution in a dynamic
environment. The model is simple and analytically tractable at a
mean-field level, and provides a general view of the gene-environment
dynamics.  Our work follows the statistical physics approach to
evolutionary dynamics,
which has become increasingly influential \cite{drossel2001biological,Sella05072005,nowak2006evolutionary,de2011contribution},
clarifying, for example, crucial issues such as the role of the topology
defining the interaction patterns in an evolving population
\cite{PhysRevLett.98.108106} or system-size effects
\cite{PhysRevLett.95.238701}. 

The model describes a large population of $n$ individuals and an
external, environmental feature. Individuals are divided in three
general types: ``specialists'' who are adapted to the environment,
``generalists'' whose fitness is independent of the environment, and ``maladapted.'' 
We assume that a complex network of genes codes for the
ability to adapt to a specific feature of the environment. For example, we might focus on the linguistic environment, which has many specific forms corresponding to particular languages, such as English. A perfect tuning of this
network would allow a ``specialist'' individual to
learn very rapidly the specific language for which it is optimized, thus
increasing her fitness (ability to survive and reproduce). Here we can
say that the genes are aligned with a specific feature of the environment
\cite{chater2009restrictions}. However, specialization comes with a
cost, reducing the flexibility of the specialized genome
\cite{ebert1998experimental}.  Even a slight environmental change
might cause problems to the offspring inheriting a genetic machinery
evolved for the original, but now different, environment. The new
individual would in fact be \textit{mis}aligned (maladapted) and her
fitness would be lowered. For this reason we include in the model also
a third kind of genomes, namely the neutrals (or ``generalists''
\cite{legros2010experimental}), for which
the fitness of an individual is independent of the specific environment. Note that working with just one
environmental feature corresponds to the assumption, standard
when modeling complex biological phenomena
\cite{baronchelli2006sharp}, that different features of the environment have roughly independent impacts on fitness.

The dynamics of the model is defined in terms of evolutionary rules
that depend on three basic parameters: The genetic mutation rate $m$,
the rate of environmental change $\ell$, and a parameter $p$,
indicating the probability that an environmental change could lead to
conditions favorable for previously maladapted individuals. As we will
see, the probability $p$ turns out to induce only small corrections in
the biological limit $p\to0$.  In terms of the remaining parameters
$m$ and $\ell$, a non-trivial mean-field phase diagram can be drawn,
exhibiting different phase transitions, akin to the so-called ``error
catastrophe'' \cite{eigenquasispecies}, as a function of $m$ for small
and large values of the rate of environmental change. This phase
diagram describes the general conditions for microevolutionary
adaptation in the presence of environmental stresses, and explains
different empirical observations of adaptation in changing
environments in a single framework.

\section*{Results}

\subsection*{Model definition}

The model is defined in terms of three different types, namely $S$,
$N$, and $M$, that represent Specialized, generalist/Neutral, and Maladapted
individuals, respectively.  Our aim is to describe a population in
which the environment changes. Thus, thinking in terms of a
theoretical fitness landscape \cite{Wright1932}, we assume that it
exhibits a maximum whose position changes whenever there is an
environmental variation, i.e., the maximum represents a \textit{moving
  target}.  The
main simplification of our model, as opposed to standard quasi-species
approximations \cite{PhysRevLett.84.191,PhysRevE.82.021904}, consists
in considering the class of specialists $S$ not as a \textit{fixed}
genome, but as the \textit{set} of those genomes which are closer to
the maximum of the fitness landscape, whatever the position of this
maximum might be.  In this theoretical fitness landscape we assume
also the presence of a secondary, local, maximum, of lower height,
representing the neutral genomes which are not affected by
environmental changes. The position of this secondary maximum is
considered static, since neutrals do not react to changes in the
environment. From this perspective, our model borrows from
quasi-species models in multiple-peaked landscapes
\cite{PhysRevE.73.041913}, with the proviso that the absolute maximum
moves in time, and we do not focus on fixed genomes, but on the set of
those close to the maxima.

\begin{figure}[h]
  \begin{center}
    \includegraphics*[width=0.44\textwidth]{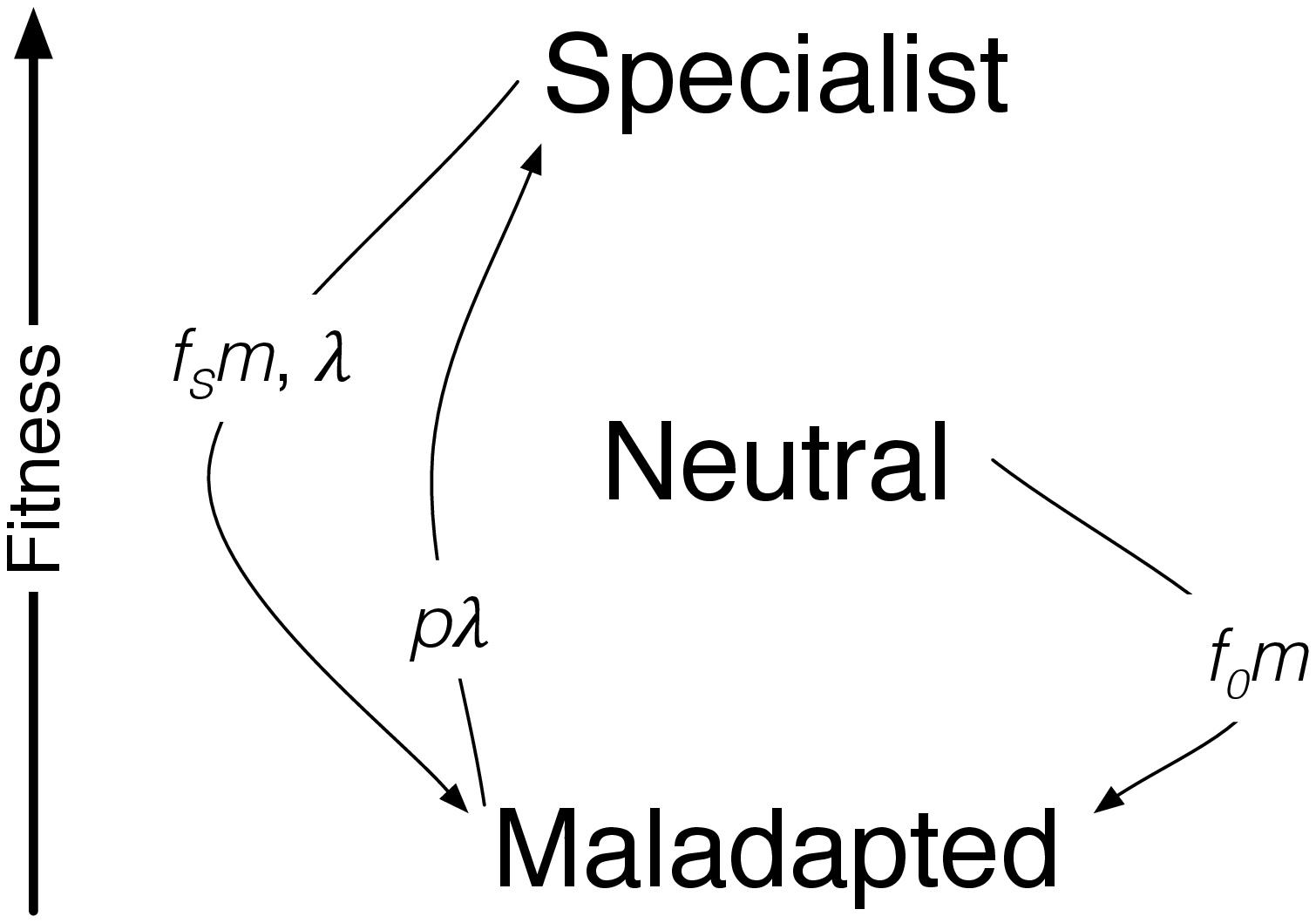}
  \end{center}
  \caption{\textbf{Schematic representation of the model definition.}
    In reproduction, the number of offspring is, of course,
    proportional to the fitness of the parent genome. Genetic
    mutations happen with probability $m$, and are detrimental, and
    leading to offspring in the maladapted class. Environmental
    changes occur with probability $\lambda$, independently from the
    state of the population. Such changes typically damage
    specialists, but also favor previously maladapted individuals. The
    latter case is however less frequent, and it is therefore
    modulated by a further probability $p$.  }
  \label{fig:modeldel}
\end{figure}

In mathematical terms, the class $S$ can thus be described as the set
of genomes that are close to the principal
maximum, by a distance $\epsilon_S$. Analogously, species $N$
represents the set of genomes close to the perfectly neutral genome
(the secondary fixed maximum), by a distance $\epsilon_N$. Finally,
the set $M$ is composed of the remaining possible genomes. We assume a
haploid reproduction system, with a fitness for each class
$f_\alpha$, satisfying the restriction $f_M < f_N < f_S$.  We consider
these fitnesses as constant, independent of the environmental
changes. At a mean-field level, assuming homogeneous mixing, the
dynamics of the model is defined as follows (see
Fig.~\ref{fig:modeldel}): Reproduction is performed by selecting an
individual with probability proportional to its fitness, as in
standard haploid models (i.e., the Moran process
\cite{moran1962spe}). The individual then produces an offspring which
is equal to itself with probability $1-m$, and that mutates to a
different type with probability $m$. Conservation of individuals is
achieved in reproduction by eliminating a randomly chosen
individual. Crucially, all genetic mutations are assumed to be
harmful, because the probability that they will lead to an increase of fitness
is negligible \cite{sawyer2007prevalence}. Therefore, a genome of
type $S$ or $N$, when mutating, reproduces into a type $M$, while $M$
genomes always reproduce into $M$ individuals.  Environmental changes
correspond to a shift of the position of the principal maximum of the
fitness landscape. This shift is assumed to take place at each time
step with a small probability $\lambda$, and produces different
effects on the three species $S$, $N$ and $M$.  Specifically, a changing environmental does not, by assumption,
affect the neutrals $N$. But the shift is mainly
unfavorable to $S$ individuals, which were best adapted to the
previous position of the maximum. This effect is implemented by
selecting, with probability $\rho_S$, a specialized individual that
will become maladapted, i.e., of class $M$. Finally, the shift could
have a beneficial effect on other previously maladapted individuals,
who were, in genomic space, far from the previous position of the
principal maximum but are now close to its present one. This effect,
which we assume to be rarer, is implemented by choosing with a small
probability $p$ a maladapted individual (with probability $\rho_M$),
which will become specialized. Note that we
  neglect backward genetic mutations from $M$ genomes to either $S$
  or $N$ species. Thus, we are considering the most common scenarios
  in which beneficial mutations are much less frequent than harmful
  ones (see for example
  \cite{eyre2007distribution,sanjuan2004distribution}).  Moreover, from
  the rules of the model, the population size $n$ is constant. This
  restriction is not problematic for our purposes, since we are
  interested only in the ratios between the population densities of
  the different species. In what follows, however, we will consider
  the limit of an infinite population, $n \rightarrow \infty$, so that
  the presence of $S$ and $M$ individuals can be assumed to be
  non-zero at the outset due to generic variability in the population,
  even thought they may be few in number. As we will see, the solution to our equations does
  not depend on the initial fractions of the different genomes.

\subsection*{Mean-field rate equations}

Let us define $\rho_\alpha$ as the density of individuals in state
$\alpha \in [S,N, M]$, satisfying the normalization condition
$\sum_\alpha \rho_\alpha = 1$. At the mean-field level, disregarding
spatial fluctuations and stochastic fluctuations, and in the
  limit of $n \rightarrow \infty$, a mathematical description of our
model can be readily obtained in terms of rate equations for the
variation of the densities $\rho_\alpha$.  To construct those, we
consider that a genome $\alpha$ increases its number (\textit{i}) when
an individual $\alpha$ is chosen for reproduction and her offspring
replaces an individual belonging to a genome $\beta \neq \alpha$,
without any mutation, i.e., with probability $(1-m)$, or (\textit{ii})
when a mutation event leads an individual with genome $\beta \neq
\alpha$ to reproduce into $\alpha$.  Conversely, a genome $\alpha$
decreases when one individual belonging to it is randomly selected for
replacement.  Additionally, $S$ genomes may decrease their number due
to a damaging change of the environment, while they can increase their
number through a (rarer) beneficial environmental change. The
corresponding rate equations take thus the form, writing explicitly
all contributions to the change to each $\rho_\alpha$,
\begin{eqnarray*}
  \dot{\rho}_S &=& \Ffac{S}(1-\rho_S)(1-m) - \Ffac{S} \rho_S m -
  \Ffac{N} \rho_S(1-m) - \Ffac{N} \rho_S m \\
  &-& \Ffac{M} \rho_S - \ell
  \rho_S + \ell p \rho_M, \nonumber \\
  \dot{\rho}_M &=& -\Ffac{S} \rho_M (1-m) + \Ffac{S} (1-\rho_M)m  -
  \Ffac{N} \rho_M (1-m) 
  + \Ffac{N} (1-\rho_M) m \\
  &+& \Ffac{M} (1-\rho_M)
  + \ell \rho_S - \ell p \rho_M,\\
  \dot{\rho}_N &=& -\Ffac{S} \rho_N (1-m) - \Ffac{S} \rho_N m +
 \Ffac{N} (1-\rho_N) (1-m) -\Ffac{N} \rho_N m - \Ffac{M} \rho_N, 
\end{eqnarray*}
where we have defined \textit{rate} of environmental change
$\ell=\lambda/(1-\lambda)$, the average fitness of the population
$\phi = \sum_\alpha f_\alpha \rho_\alpha$, and we have performed an
irrelevant rescaling of units of time.  After some algebraic
manipulations, the previous equations can be simplified to the form:
\begin{subequations}
\begin{eqnarray}
  \dot{\rho}_S &=& \rho_S \left[ -1 + (1-m) \hf{S}
    -\ell\right] + p \ell \rho_M,\label{eq:1}\\ 
  \dot{\rho}_N &=& \rho_N \left[ -1 + (1-m) \hf{N}\right],\label{eq:2}\\
  \dot{\rho}_M &=& \rho_M \left[ -1 + (1-m) \hf{M}  -p \ell \right] + m +
  \ell \rho_S,\label{eq:5}
\end{eqnarray}
\label{eq:3}  
\end{subequations}
where we have defined $\hf{\alpha} = f_\alpha/\phi$.  

Equations \eqref{eq:3} completely define the dynamics of the model at
the mean field level. In the following we will analyze
their solution in different limits.

\subsection*{Analytical solution}

In the long term steady state, the solutions of this dynamical system
are obtained by imposing the conditions $\dot{\rho}_S = \dot{\rho}_N =
\dot{\rho}_M=0$ on Eqs.~\eqref{eq:3}, solving the ensuing algebraic
equations, and checking for the stability of the solutions, by looking
at the eigenvalues of the Jacobian matrix, evaluated at the respective
solution.  The solutions obtained this way in the general case $p>0$
turn out to be quite complex, so in order to simplify the resulting
expressions, we choose particular values of the fitnesses, namely $f_S
= 2$, $f_N = 1$ and $f_M=0.5$, respecting their natural
ordering. Solutions for other values can be obtained using the same
steps.

\subsubsection*{Case $p \neq 0$}

The algebraic equations ruling the steady state, obtained from
Eqs.~\eqref{eq:3} by setting to zero the time derivatives, can be
solved using a standard computational software package. This results
in three sets of solutions, taking the form
\begin{eqnarray*}
  \mathcal{S}_1 &&
  \begin{cases}
    \rho_N = 0\\
    \rho_S = -\dfrac{\sqrt{2 \ell (m (4-8 p)+12 p-3)+(3-4 m)^2+(4 p
        \ell +\ell )^2}+4 m-2
      p \ell +\ell -3}{6 (p \ell +\ell +1)} \\
    \rho_M= \dfrac{\sqrt{2 \ell (m (4-8 p)+12 p-3)+(3-4 m)^2+(4 p
        \ell +\ell )^2}+4 m+4 p \ell +7 \ell +3}{6 (p \ell
      +\ell +1)}
  \end{cases}, \\
 \mathcal{ S}_2 &&
 \begin{cases}
   \rho_N = 0\\
   \rho_S = \dfrac{\sqrt{2 \ell (m (4-8 p)+12 p-3)+(3-4 m)^2+(4 p
       \ell +\ell )^2}-4 m+2 p
     \ell -\ell +3}{6 (p \ell +\ell +1)}\\
   \rho_M = \dfrac{-\sqrt{2 \ell (m (4-8 p)+12 p-3)+(3-4 m)^2+(4 p
       \ell +\ell )^2}+4 m+4 p \ell +7 \ell +3}{6 (p \ell
     +\ell +1)}
 \end{cases}, \\
 \mathcal{ S}_3 &&
 \begin{cases}
   \rho_N = \dfrac{2 \ell (m p+m+p)-2 m-\ell +1}{(2 p-1) \ell
     +1}\\
   \rho_S = -\dfrac{2 m p \ell }{2 p \ell -\ell +1}\\
   \rho_M = -\dfrac{2 m (\ell -1)}{(2 p-1) \ell +1}
 \end{cases}.
\end{eqnarray*} 

Solution $\mathcal{S}_1$ describes a $\rho_S$ density that is negative 
in the parameter region $p>0$, i.e., it is an ``unphysical'' solution that does not
describe any realistic scenario. Solution $\mathcal{S}_2$ has 
nonzero densities in the whole parameter space, while solution $\mathcal{S}_3$ 
is physical only in the region
\begin{equation}
  \label{eq:9}
  \mathcal{R }= \left\{0<m<\frac{1}{2}\land 
    0<p<\frac{1-2 m}{2 m+2}\land 
    \ell \geq \frac{2 m-1}{2 (m p+ m+ p)-1} \right\} . 
\end{equation}
In order to find the relative stability of the physical solutions
$\mathcal{S}_2$ and $\mathcal{S}_3$, we consider the Jacobian matrix
of the equation system Eqs.~\eqref{eq:3}, taking the form
\begin{equation*}
  \label{eq:10}
 J = \left(
\begin{array}{ccc}
 -\frac{4 (m-1) \left(\rho _M+2 \rho _N\right)}{\left(\rho _M+2 \rho _N+4 \rho
   _S\right){}^2}-\ell -1 & \frac{8 (m-1) \rho _S}{\left(\rho _M+2 \rho _N+4 \rho
   _S\right){}^2} & \frac{4 (m-1) \rho _S}{\left(\rho _M+2 \rho _N+4 \rho
   _S\right){}^2}+p \ell  \\
 \frac{8 (m-1) \rho _N}{\left(\rho _M+2 \rho _N+4 \rho _S\right){}^2} & -\frac{2
   (m-1) \left(\rho _M+4 \rho _S\right)}{\left(\rho _M+2 \rho _N+4 \rho
   _S\right){}^2}-1 & \frac{2 (m-1) \rho _N}{\left(\rho _M+2 \rho _N+4 \rho
   _S\right){}^2} \\
 \frac{4 (m-1) \rho _M}{\left(\rho _M+2 \rho _N+4 \rho _S\right){}^2}+\ell  &
   \frac{2 (m-1) \rho _M}{\left(\rho _M+2 \rho _N+4 \rho _S\right){}^2} & -\frac{2
   (m-1) \left(\rho _N+2 \rho _S\right)}{\left(\rho _M+2 \rho _N+4 \rho
   _S\right){}^2}-p \ell -1
\end{array}
\right).
\end{equation*}
We then compute the eigenvalues of matrix $J$, evaluated for the
different solutions $\mathcal{S}_2$ and $\mathcal{S}_3$. In any given
region of parameter space, the stable solution (in the stationary
limit) is the one possessing a negative largest
eigenvalue. Examination of these eigenvalues, operation performed again
with the help of standard computational software packages, leads to
the solution:

$\bullet$ If $(m,p, \ell) \in \mathcal{R}$:
\begin{equation}
  \begin{cases}
    \rho_N = \dfrac{2 \ell  (m p+m+p)-2 m-\ell +1}{(2 p-1) \ell
      +1} \\
    \rho_S = \dfrac{2 m p \ell }{\ell -2 p \ell -1}\\
    \rho_M = \dfrac{2 m (\ell -1)}{(1-2 p) \ell +1}
  \end{cases}.
  \label{eq:16}
\end{equation}

$\bullet$ Otherwise:
\begin{equation}
  \label{eq:17}
  \begin{cases}
    \rho_N = 0\\
    \rho_S =\dfrac{\sqrt{2 \ell  (m (4-8 p)+12 p-3)+(3-4 m)^2+(4 p
        \ell +\ell )^2}-4 m+2 p \ell -\ell +3}{6 (p \ell 
      +\ell +1)} \\
    \rho_M = \dfrac{-\sqrt{2 \ell  (m (4-8 p)+12 p-3)+(3-4 m)^2+(4
        p \ell +\ell )^2}+4 m+4 p \ell +7 \ell +3}{6 (p
      \ell 
      +\ell +1)}
  \end{cases}
\end{equation}
where $\mathcal{R}$ is the domain in the parameter space defined in
Eq.~\eqref{eq:9}.

\begin{figure}[h]
  \begin{center}
    \includegraphics*[width=0.95\textwidth]{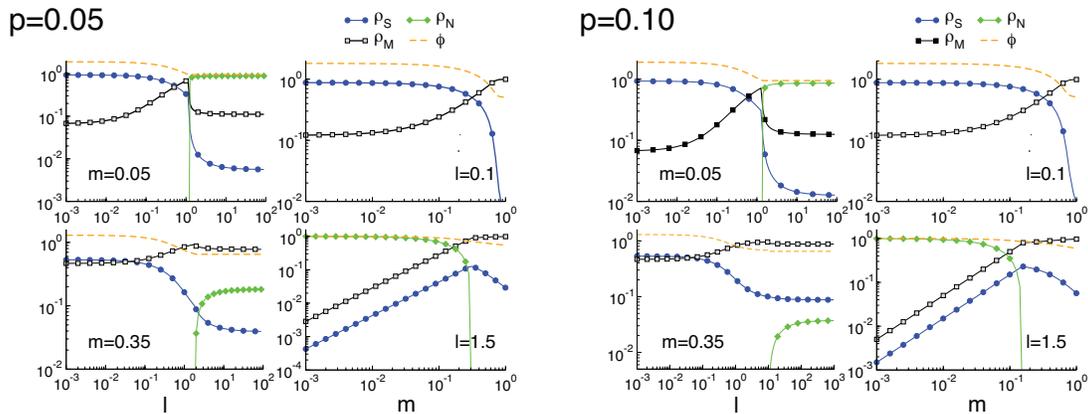}
  \end{center}
  \caption{\textbf{Species densities at the steady state for small
      $p$}. Densities of $S$ (blue), $N$ (green) and $M$ (gray)
    genomes as a function of the environmental mutation rate $\ell$
    for fixed $m$, and as a function of $m$ for fixed $\ell$. The left
    panel corresponds to $p=0.05$, and the right panel to
    $p=0.10$. Dashed orange lines represent the average fitness of the
    population.}
  \label{fig:slicesnonzerop}
\end{figure}

\begin{figure}[h]
   \begin{center}
     \includegraphics*[width=0.95\textwidth]{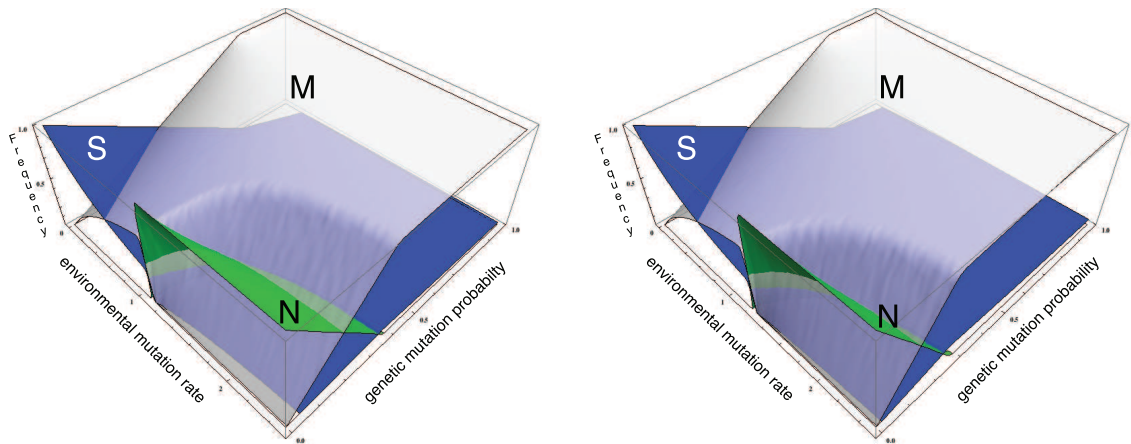}
   \end{center}
  \caption{\textbf{Species densities at the steady state for small
      $p$}.  Densities of $S$ (blue), $N$ (green) and $M$ (gray)
    genomes as a function of the environmental mutation rate $\ell$
    and the genetic mutation rate $m$, for fixed values $p=0.05$
    (left) and $p=0.1$ (right). Fitness values are $f_S = 2$, $f_N =
    1$ and $f_M=0.5$. }
  \label{fig:sim3dnonzerop}
\end{figure}

The analytical solutions given by Eqs.~\eqref{eq:16} and~\eqref{eq:17}
are quite complex, and it is difficult to extract direct
interpretations from them.  However, the behavior of the solutions can
be understood in the biologically relevant region of small $p$.
Fig.~\ref{fig:slicesnonzerop} shows the densities $\rho_\alpha$ as a
function of $m$ or $\ell$, at fixed $\ell$ and $m$ respectively, for
two values of $p$, along with the value of the total fitness
$\phi=\sum_\alpha \rho_\alpha f_\alpha$.  In the upper left corner
plot for each value of $p$, we consider the case $m=0.05$,
representative of the biologically relevant scenario of small mutation
rate.  When $\ell$ is very small, most of the population stays aligned
with the environmental feature and $\rho_S \sim 1$. As $\ell$ increases,
maladapted genomes appear and eventually overcome the specialized
genes. At a definite value of $\ell$, however, a discontinuity takes
place and neutral individuals suddenly appear and become the majority
of the population, while the density of both maladapted and
specialists decreases. The decrease in specialists is larger for
larger values of $p$. Thus, for sufficiently large $\ell$ and small
$m$, trying to catch up with the rapidly evolving environmental feature
is not a viable strategy, since the risk of producing a
maladapted offspring becomes destructive. Interestingly the strategy
adopted by the majority of the individuals guarantees the maximum
average fitness in any given region of the $(\ell, m)$ plane, for
every value of $p$. For large values of $m$ (lower left plots), the
situation is qualitatively different. For small $\ell$, maladapted and
neutral individuals are almost equally numerous. When $\ell$ increases beyond a
threshold, neutral individuals again appear suddenly, but they are
unable to overcome maladapted genomes. Only for large values of $p$
are neutrals capable to prevail over the specialists.  The right plots
for each value of $p$ in Fig.~\ref{fig:slicesnonzerop} show the
evolution of the species' densities as a function of $m$ for fixed
$\ell$. In this case, for small $\ell$ neutrals are absent from the
system, and there is a simple competition between specialists and
maladapted, the former being predominant for small genetic mutation
rates, but going extinct for large $m$. On the other hand, when the
rate of environmental change is sufficiently large, we enter a new
scenario in which neutrals are predominant for small $m$. Beyond a
mutation rate threshold, however, neutrals suddenly become extinct, and
their population is replaced by maladapted genomes, while specialists
decrease their density for large $m$. Interestingly, in this region of
large $\ell$, specialists can survive even for very large mutation
rates, close to $1$, due to the effect of a nonzero $p$ that prevents
their complete elimination. Fig.~\ref{fig:sim3dnonzerop}
shows the complete picture of the relative species' abundances as as a
function of $m$ and $\ell$ for the previously considered values of
$p$.

\subsubsection*{Case $p=0$}
\label{sec:case-p=0}

A more precise mathematical characterization of the phenomenology
discussed above, and in particular of position of the transitions
taking place for different values of $m$ and $\ell$, can be obtained
in the particular case $p=0$. Here, qualitative arguments allow us to
solve the model in a much simpler way, for general values of $f_S$,
$f_N$ and $f_M$. This analysis, moreover, reveals the role of
$p$ in the dynamics of our model.

The relevant equations in the $p=0$ case read
\begin{subequations}
\begin{eqnarray}
  \dot{\rho}_S &=& \rho_S \left[ -1 + (1-m) \hf{S}
    -\ell\right], \label{eq:11}\\ 
  \dot{\rho}_N &=& \rho_N \left[ -1 + (1-m) \hf{N}\right],\label{eq:12}\\
  \dot{\rho}_M &=& \rho_M \left[ -1 + (1-m) \hf{M}  \right] + m +
  \ell \rho_S.\label{eq:15}
\end{eqnarray}
\label{eq:13}  
\end{subequations}
To find their solution, we argue as follows: If $m$ is very close to $1$,
all terms in square brakets in Eqs.~\eqref{eq:13} will be
negative. Therefore, the only stable solution will be $\rho_S =
\rho_N=0$, $\rho_M = 1$, for any value of $\ell$. Under this
conditions, the quantities within square brackets in
Eqs.~(\ref{eq:11}) and (\ref{eq:12}) take the form $-1 -\ell
+(1-m)f_S/f_M$ and $-1 +(1-m)f_N/f_M$, respectively. Decreasing the
value of $m$, the first solution with $\rho_M <1$ will take place for
the first of these values that become zero. This occurs when $m$ is smaller
than either $m_{c,1}(\ell) = 1-(\ell+1)f_M/f_S$ or $m_{c,2} =
1-f_M/f_N$, respectively. Since $f_M < f_N$, we have $0 < m_{c,2} <1$,
and this transition will always be physical. However, for $\ell >
\ell_{c,1} = (f_S - f_M)/f_M$, we have $m_{c,1}(\ell) <0$ and it is
not physical. In this case, when $\ell > \ell_{c,1}$, if $m <m_{c,2}$,
$\rho_S$ decays exponentially, and in the long time limit $\rho_S=0$;
the existence of a non-zero $\rho_N$ solution imposes
$-1+(1-m)\hf{N}=0$, or $\phi = (1-m)f_N$, from where the restricted
normalization condition $\rho_N + \rho_M=1$ leads to the solution
$\rho_N = 1 - m/m_{c,2}$ $\rho_M = m/m_{c,2}$, and $\rho_S=0$. In the
case $\ell < \ell_{c,1}$, which density $\rho_S$ or $\rho_N$ becomes
first non-zero depends on which threshold, $m_{c,1}(\ell)$ or
$m_{c,2}$ is larger. Thus, if $\ell > \ell_{c,2} = (f_S - f_N)/f_N$,
then $m_{c,2} > m_{c,1}(\ell)$. Therefore, when decreasing $m$, the
first density to take a non-zero value is $\rho_N$. $\rho_S$ decays
again exponentially, so the solution is the same as in the case $\ell
\geq \ell_{c,1}$. Finally, for $\ell < \ell_{c,2}$, $\rho_S$ is the
first density to become non-zero when $m<m_{c,1}$. The steady state
solution comes from imposing $-1-\ell +(1-m)\hf{S}=0$, leading to
$\phi = f_S (1-m)/(\ell+1)$. In this region, the factor in square 
brackets in Eq.~(\ref{eq:12}) becomes negative, indicating an
exponential decay and a corresponding steady state value
$\rho_N=0$. We are therefore led to the solution, using the
normalization condition, $\rho_S =
(m_{c,1}(\ell)-m)/[\ell_{c,1}(1-m_{c,1}(\ell))]$, $\rho_M =1-\rho_S$.

The final solution in this case can thus be summarized as follows:
\begin{itemize}

\item  For $\ell < \ell_{c,2} = (f_S - f_N)/f_N$
  \begin{itemize}
  \item If $m < m_{c,1}(\ell) =  1-(\ell+1)f_M/f_S$
    \begin{equation}
      \rho_S  =  \frac{f_M}{f_S-f_M}
      \left[\frac{m_{c,1}(\ell)-m}{1-m_{c,1}(\ell)} \right], \quad
      \rho_M = 
     1- \rho_S, \quad \rho_N=0. \label{eq:14}
    \end{equation}
  \item If $m \geq  m_{c,1}(\ell)$
    \begin{equation}
      \rho_S  =  \rho_N=0, \qquad \rho_M = 1 .
    \end{equation}
  \end{itemize}

\item For $\ell \geq \ell_{c,2} =(f_S - f_N)/f_N$:
  \begin{itemize}
  \item If $m < m_{c,2} =  1-f_M/f_N$
    \begin{equation}
      \rho_N  =  1 - \frac{m}{m_{c,2}}, \quad \rho_M =
      \frac{m}{m_{c,2}}, \quad \rho_S=0. 
    \end{equation}
  \item   If $m \geq  m_{c,2}$
    \begin{equation}
      \rho_S  =  \rho_N=0, \qquad \rho_M = 1 .
    \end{equation}
  \end{itemize}
  
\end{itemize}

\begin{figure}[h]
   \begin{center}
     \includegraphics*[width=0.44\textwidth]{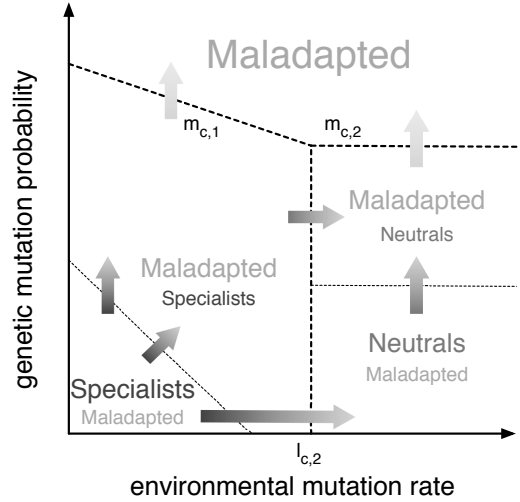}
   \end{center}
  \caption{\textbf{Phase diagram for the case $p=0$.} The most
    abundant genome is indicated by a larger name in each region. The
    average fitness of the population decreases along the
    arrows. }
  \label{fig:quadrants}
\end{figure}

Fig.~\ref{fig:quadrants} sketches the phase diagram, as a function of
$m$ and $\ell$, resulting from the previous equations. The different scenarios for small
and large values of $\ell$
are now explicit.  For small $\ell< \ell_{c,2}$, specialist
individuals (in the $N$ class) are able to survive, and even dominate
the population, as long as the mutation rate is small. In fact, for
$m<m_{c,3}(\ell) = m_{c,1}(\ell)-\ell_{c,1}[1-m_{c,1}(\ell)]/2$, the
density of specialists is larger than the density of maladapted
individuals. For larger $m$, the density of specialized genomes
decreases, until it reaches the $\ell$-dependent threshold
$m_{c,1}(\ell)$, leading to a continuous, second order, phase
transition (akin to the error catastrophe in quasispecies models
\cite{eigenquasispecies}) beyond which the whole population becomes
maladapted and thus prone to eventual extinction. In all of this
region of small $\ell$, neutral individuals are irrelevant. For small
$m$, specialists perform much better, while for large $m$ only
maladapted individuals survive.

When $\ell$ increases, a different picture emerges. For fixed, small
$m<m_{c,2}$, the explicit behavior of $\rho_S$ as a function of
$\ell$, reads we can obtain from Eq.~\eqref{eq:14}

\begin{equation}
  \label{eq:18}
  \rho_S = \frac{f_M}{(\ell+1)(f_S-f_N)}
  \left[\frac{f_S}{f_M}(1-m)-1 -\ell \right], 
\end{equation}
for $\ell<\ell_{c,2}$, and zero otherwise. Thus, when crossing
$\ell_{c,2}$, the density of specialized individuals experiences a
first order transition to extinction, with a jump of magnitude
\begin{equation}
  \label{eq:19}
  \Delta \rho_S = \frac{f_N(1-m) -f_M}{f_S-f_M}.
\end{equation}
The sudden extinction of specialized individuals coincides with the
abrupt emergence of neutrals, in a related first order transition for
$\rho_N$ with an associated jump
\begin{equation}
  \label{eq:20}
  \Delta \rho_N = \frac{f_N(1-m)-f_M}{f_N-f_M}.
\end{equation}
In this large $\ell$ region, neutrals are able to cope with
environmental change if the mutation rate is sufficiently small, again
up to a maximum mutation rate $m_{c,1}$, after which $\rho_N$
continuously becomes zero and only maladapted individuals can
survive. These discontinuous transitions as a function of $\ell$ at
fixed $m$ are reminiscent of the phenomenology observed in
quasispecies models with higher order replication mechanisms
\cite{PhysRevLett.104.188101}. We note, however, that transition as a
function of $m$ at fixed $\ell$ are all continuous.

\begin{figure}[h]
   \begin{center}
     \includegraphics*[width=0.95\textwidth]{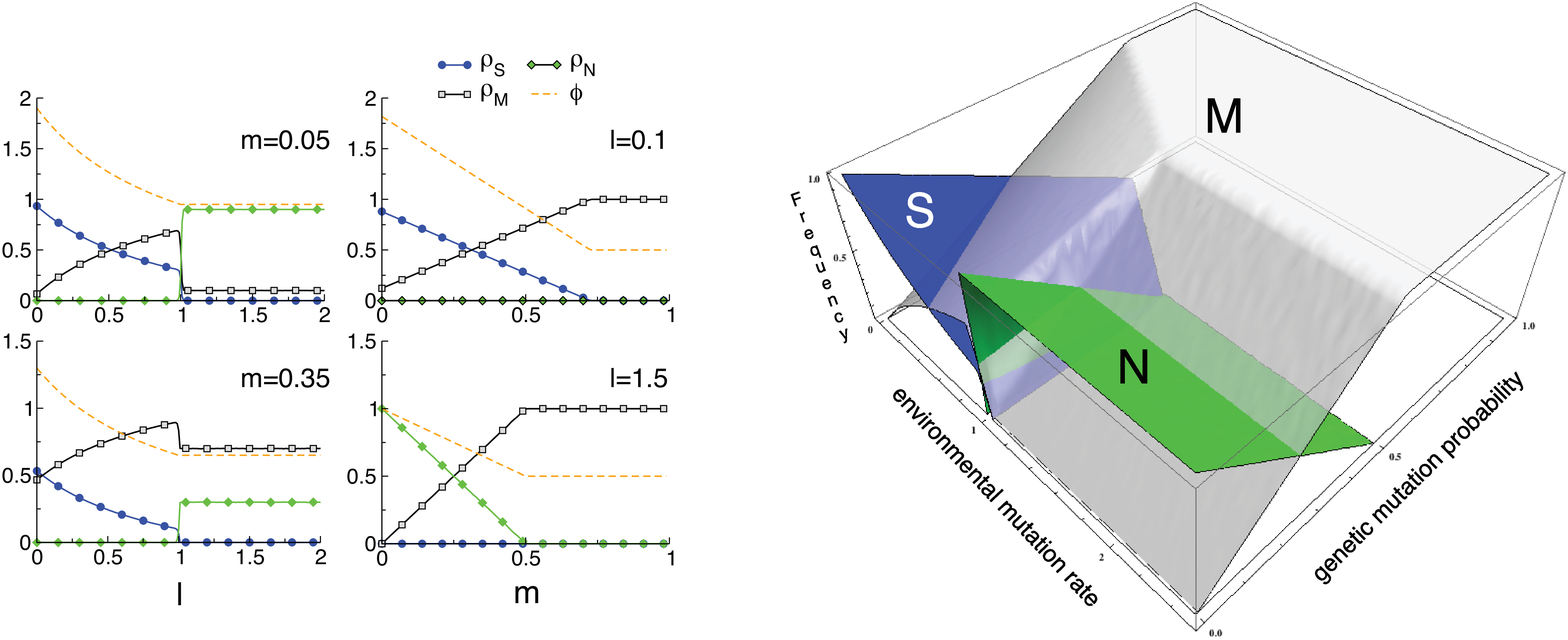}
   \end{center}
  \caption{\textbf{Species densities at the steady state for small
      $p$}. Left panel: Genome densities as a function of the
    environmental mutation rate $\ell$ for fixed $m$, and as a
    function of $m$ for fixed $\ell$.  Fitness values are chosen as $f_M
    = 2$, $f_N = 1$ and $f_M=0.5$. Hence, $\ell_{c,2}=1$,
    $m_{c,s}=0.5$ and $m_{c,1}=(1-\ell)/2$, implying $ 0.5 \leq \phi
    \leq 2$ (see main text). Right panel: Densities of $S$ (blue), $N$
    (green) and $M$ (gray) genomes as a function of the environmental
    mutation rate $\ell$ and the genetic mutation rate $m$.}
  \label{fig:sim3d}
\end{figure}

Fig.~\ref{fig:sim3d} shows the proportions of the three genomes along
with the average fitness as function of $m$ for fixed $\ell$, and as a
function of $\ell$ at fixed $m$ (left panel), and the general scenario
as a function of both $m$ and $\ell$ (right panel).  As it is clear,
while an abrupt transition occurs at the level of the genome
frequencies at $\ell_{c,2}$ ($\ell_{c,2}=1$ in Fig.~\ref{fig:sim3d}),
the average fitness $\phi$ exhibits a continuous behavior, decreasing
monotonously from the maximum $\phi=f_S$ for $m=\ell=0$ to $\phi=f_M$
for large values of $m$ (when all the other genomes simply mutate into
$M$). When $m$ is fixed (left column, left panel), increasing $\ell$
causes an increase of $M$ genomes and a simultaneous decrease in
$\rho_S$ and $\phi$. As $\ell=\ell_{c,2}$, however $S$ genomes
disappear as the neutral genomes abruptly appear. The latter
guarantees a constant value of $\phi$. $M$ genomes are constantly
created due to the genetic mutation rate, but their fitness is lower
so they do not reproduce frequently. This scenario is stable, and any
further increase in $l$ does not produce any effect. The role of $m$
is better understood at fixed values of $\ell$. When $\ell<\ell_{c,2}$
(top panel) increasing $m$ deteriorates the fitness of the population
since $S$ genomes are substituted by $M$ ones, which eventually become
fixed ($\rho_M=1$ as $m=m_{c,1}=0.725$ in figure). A similar behavior
is observed for $\ell>\ell_{c,2}$ (bottom panel), but here the $M$
genomes take the place of the $N$ genomes, till the latter disappear at
$m_{c,2}=0.5$ for the values of the simulations.

The crucial difference between the cases $p\neq0$ and $p=0$, as can be
observed from the comparison of Figs.~\ref{fig:slicesnonzerop}
and~\ref{fig:sim3d} is the effect of a positive density of
specialists for large $m$ and $\ell$. In the case $p=0$, the density
$\rho_S$ goes to zero after the corresponding transition, specialist
being unable to cope with extreme genetic and/or environmental rates
of change. In the presence of a non-zero $p$, signaling the
possibility of collateral beneficial effects of an environmental
change to previously maladapted individuals, susceptible individuals
are still able to thrive in an situation combining both fast genetic
and environmental change (see lower right plots in
Fig.~\ref{fig:slicesnonzerop}). This effect is due to the
feedback mechanism induced by the parameter $p$, that allows the
replenishment of the $S$ individuals from previously maladapted
individuals. Their prevalence is however relatively small, and
comparatively negligible with respect to the predominant species,
either $N$ or $M$, especially in the case of small populations. Finally,
it is worth noting that, while the prevalence of $M$ genomes is stable in
our model, it can be interpreted as a metastable state leading to extinction 
in a multi-species scenario.

\section*{Discussion}

The model presented in this paper shows that a genetic adaptation to a
specific form of an environmental feature is profitable only as long as
the rate of change in the environment is not too fast. Indeed, a phase
transition determines the onset of a different regime in which a neutral
strategy is advantageous. The critical value of the environmental rate
separating the two phases is proportional to the difference between
the fitness of the neutral and specialist individuals.
This analysis is the consequence of the
simplicity of the model that, in contrast to previous modeling
attempts, allows us not only to outline a qualitative scenario, but
also to characterize quantitatively and in a transparent way the role
of the different parameters, in the hope that future experimental work
will be able to test these findings.

The analysis proposed here provides a framework for understanding a range of empirical finding. 
As mentioned above, for example, lactose tolerance did
become genetically encoded \cite{beja2003gene} while language
is a moving target for the genes, that appears to change too fast to allow
genetic adaptation \cite{christiansen2008language}. In the same way,
agricultural practices that determine an increased presence of malaria are linked
to genetic mutations that cause malaria reduction \cite{wiesenfeld1967sickle, saunders2002nucleotide, laland2010culture}, and bioinformatic methods have recently shown that climate has been an important selective pressure acting on candidate genes for common metabolic disorders \cite{hancock2008adaptations}.
Another particularly significant example comes from biology, where the diversity and temporal variability 
of a population of hosts
determines the pressure for parasites to specialize on one host or to
become generalists on a wide range of hosts
\cite{crill2000evolutionary}, as it has been experimentally shown for
example in parasites \textit{Brachiola algerae} infecting
\textit{Aedes aegypti} mosquitoes \cite{legros2010experimental}. Our
model coherently predicts also that specialized genomes would decrease
their fitness if the mutation rate of the corresponding environmental
feature increases (Fig.~\ref{fig:sim3d}). Interestingly, this is what
has been observed in relation to climate change, the consequence being
a diminished robustness against competitors and natural enemies,
which, in a multi-species scenario, could eventually lead to
extinction \cite{thomas2004extinction}.

In summary, we have introduced an evolutionary model that captures
a wide array of natural scenarios in which genes evolve against a potentially changing environment. 
These results have been obtained using strong simplifying assumptions
that can be relaxed in future work.  For example, a
natural extension of the model could consider a more complex network
of environment-gene interactions, including
the possibility of feedback between genes and the
rate of change in the environment. Such a generalization could lead to
important results and a richer phenomenology
\cite{lande1983measurement,kirkpatrick2009patterns}, as well as
enlarge the range of applicability of the model \cite{Gilman:2012fk},
even though it may reduce the mathematical tractability of the
resulting equations. Likewise, the fitness of each genome could depend,
for example, on its relative abundance in the population, instead of
being a constant parameter. Finally, the equations we have derived 
apply in the case of very large populations;  a possible extension could consider the 
effects of fluctuations in small groups.
The framework we have put forth is general
and allows these and other aspects (such as the effects of spatial
fluctuations in finite dimensions) to be addressed in a principled
way.




\clearpage
\section*{Figures}

\vspace{0.5cm}

\end{document}